\documentclass[aps,prl,twocolumn,groupedaddress,showpacs]{revtex4}

\usepackage{epsfig}

\begin{document}

\title{Spontaneous emergence of van der Waals interaction in piezo-resonators   - a road to phase coherence at mK temperatures}

\author{M. \v{C}love\v{c}ko}
\author{P. Skyba}
\email[Electronic Address: ]{skyba@saske.sk}
\author{F. Vavrek}

\affiliation{{\mbox Centre of Low Temperature Physics, Institute  of Experimental
Physics, SAS} and P. J. \v{S}af\'arik University Ko\v{s}ice, Watsonova 47, 04001
Ko\v{s}ice, Slovakia.}

\date{\today}

\begin{abstract}

We present the experimental results on the spontaneous emergence of the phase coherence
in the system of oscillating electric dipoles in quartz piezo-resonators caused by the
van der Waals interaction. Spontaneous emergence of the phase coherence in these systems
is manifested via temperature-dependent, extremely accurate tune-up of their resonance
frequencies in 9th order with relative spectral line-width $\delta f_0/f_0$ less than
3.10$^{-8}$ (this number is comparable with that in lasers) along with the very high
frequency stability characterized by the low values of the Allan deviations. We also show
that the application of an incoherent (noise) excitation signal leads to a spontaneous
formation of the phase coherent state, and that the dissipation processes do not affect
this phase coherent state (i.e. the resonance frequency of the system). All
above-mentioned signatures are typical characteristics for a Bose-Einstein condensate of
excitations. Smallness of the relative spectral lime-width in quartz piezo-resonators
opens their potential application as alternative time etalons.

\end{abstract}


\maketitle

Bose-Einstein (B-E) condensation is a fundamental physical phenomenon when a macroscopic
number of bosons condense into a collective quantum ground state governed by a single
wave function. Text-book examples of the B-E condensates are superfluid $^4$He and
ultracold atomic gases \cite{kapitza,ketterle,cornell}. In case of the fermions, the
scenario of the B-E condensation is provided via a mutual coupling interaction between
pair of fermions allowing them the formation of the bosonic Cooper pairs, and these pairs
condense and occupy the single energy level that is lower than the Fermi energy by the
energy gap $\Delta$ . Examples of such B-E condensates are superconductivity of electrons
and superfluid $^3$He \cite{bcs,osheroff}.  Recently, however, a concept of the B-E
condensation was extended also to the physical systems with a spontaneous emergence of
the phase coherence \cite{snoke1}. These can be the systems of excitations, the lifetime
of which is longer than the time they need to scatter and there is an interaction acting
between them allowing to set a single energy state governed by the single wave function
\cite{snoke2,stoof}. Typical examples of such excitations involve magnons, excitons,
polaritons, etc. \cite{demo,magnons,kasp,exci1,exci2,szym}. It is worth to note that
phenomenon of the spontaneous emergence of the order (or a synchronization process) is
considered to be a vast and more general process acting in many complex systems
\cite{report}.

In physical systems, the criterium of the spontaneous emergence of the phase coherence
means that particular system under consideration should exhibit a weak long-range
interaction being comparable with its thermal energy, and by cooling the system, this
interaction (i) should overwhelm the thermal energy and (ii) should couple and tune the
excitations in a phase-coherent state. The dipole-dipole interactions, and in particular
the van der Waals interactions are one of the fundamental, but relatively weak
interactions which could satisfy above mentioned criteria and they are presented in many
physical systems.

We focused our investigation on the presence of the van der Waals interactions in
piezoelectric resonators. We used commercially available quartz tuning forks resonating
at nominal frequencies 32\,kHz, 77\,kHz and 100\,kHz. These devices are intensively used
in AFM and STM techniques \cite{afm1,afm2,stm}, physics of superfluids ($^3$He and
$^4$He) \cite{rob,blaz1,hel1}, etc.  It is generally known that when the standard quartz
tuning forks oscillating in vacuum are cooled down to temperatures below $\sim$ 15\,K,
these resonators undergo a "thermodynamic" transition into the high Q-value oscillation
mode with the Q-value of the order of 10$^6$ or more \cite{rob,fork1}. In order to
measure quartz tuning fork with high Q-value, instead of using a traditional technique of
the frequency sweep with continuous voltage excitation, we adapted and applied a
pulsed-demodulation (P-D) technique (or heterodyning technique with pulsed excitation)
\cite{fork2}. The P-D technique transforms the frequency of the measured signal to lower
values without losing information about the original frequency and allows reduction of
the sampling rate and increases the resolution of the frequency measurements. This allows
to measure the decay signals from the freely oscillating resonators with high resolution
in the frequency in 9th order. Free oscillations of the tuning forks at resonance
frequency $f_{tf}$ generate an alternating piezoelectric current being detected by a
custom-made current-to-voltage (I/V) converter in order to minimize the losses in
detection circuit \cite{IV,JLTP}.  The I/V converter's output voltage signal is then
measured by a lock-in amplifier operating as a demodulator.

\begin{figure}[htb]
%\begin{center}
\epsfig{figure=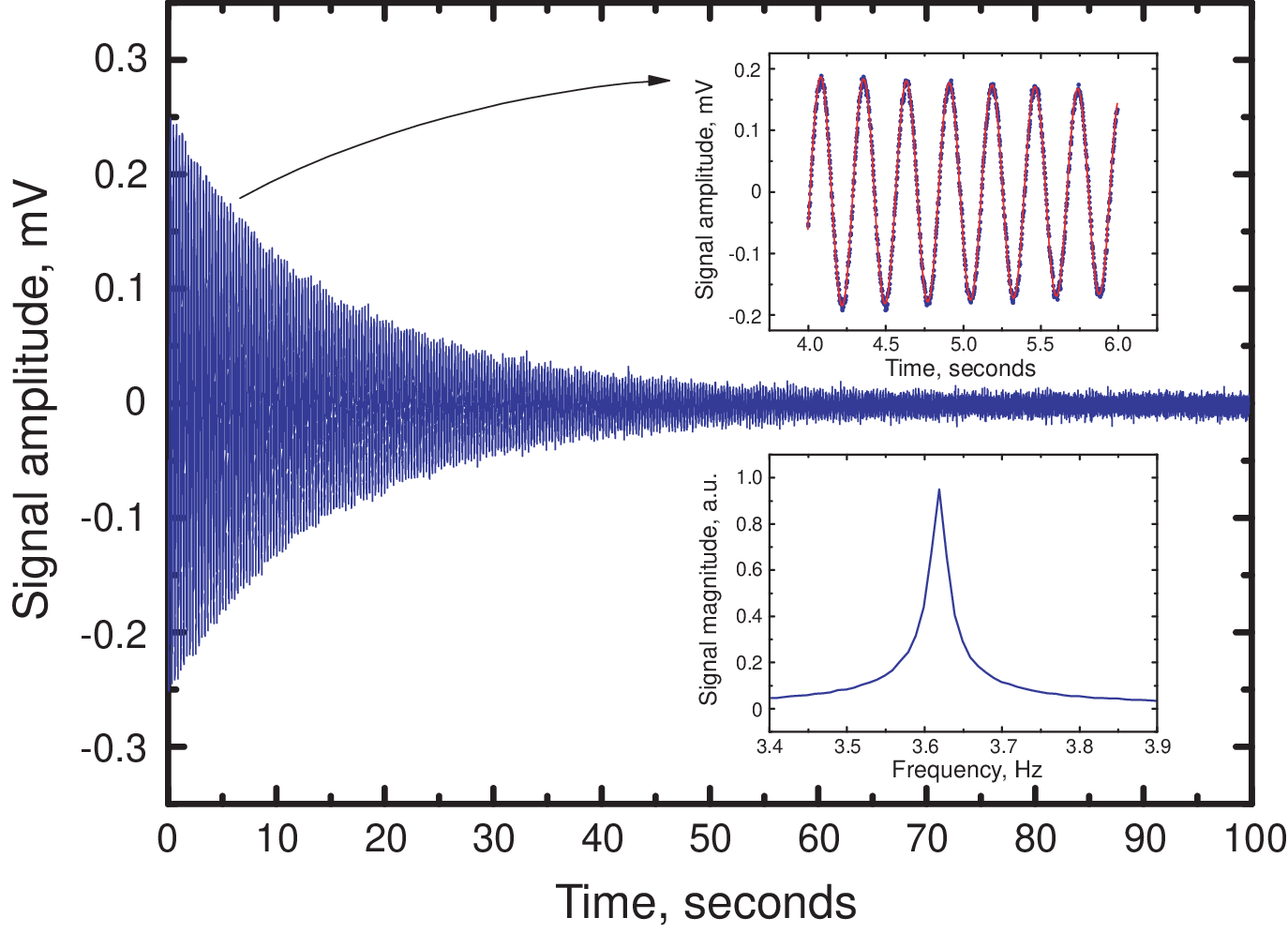,width=75mm} \caption{(Colour online) Free decay signal of the 32\,kHz
tuning fork's oscillations measured using pulsed-demodulation technique. Insets: a time window
showing fit (red line) to experimental data (points) using equation (\ref{equ1}) (top),
FFT of the presented signal showing the precision of the resonance frequency determination
(bottom).} \label{fig1}
%\end{center}
\end{figure}

Before their installation the tuning forks were removed from the original metal can and
the former (magnetic) leads were replaced by copper wires having a diameter of
120\,$\mu$m. These copper wires were electrically connected to tuning fork's pads using a
conductive silver epoxy and glued to a small piece of Stycast 1266 - epoxy impregnated
paper in order to achieve a mechanical stiffness of the set-up. Cooling of the tuning
forks was ensured by clamping these copper wires between two copper blocks and these
blocks were screwed to the mixing chamber of the cryogen-free dilution refrigerator
Triton 200. As an example, Fig. \ref{fig1}  shows the  decay signal of 32\,kHz tuning
fork's free oscillations measured in vacuum at temperature of 25\,mK. The deposited
energy (250 pulses with 5\,mV$_{RMS}$ amplitude) used to excite the fork is equal to
33$\cdot$10$^{-15}$\,Joule. The tuning fork's high Q-value is demonstrated by a long
lasting decay signal of the order of several tens of seconds. Decay signals measured from
77\,kHz and 100\,kHz tuning forks had the same form, however, the duration of their decay
signals in time was slightly shorter, up to 30\,seconds for both  77\,kHz and 100\,kHz
tuning forks. All measured decay signals were fitted using the expression
\begin{equation}
V(t)=V(0)\,e^{-\frac{t}{\tau}}\sin(2\pi f_{sig} t + \phi),\label{equ1}
\end{equation}
where $V(0)$ is the initial amplitude of the signal, $\tau$ is the relaxation time
constant characterizing a damping process, $f_{sig}$ is the signal frequency and $\phi$
is the signal phase. Insets to Fig. \ref{fig1} show a time window with the fit to
experimental data using equation (\ref{equ1}) and FFT spectrum of the signal. Fit applied
to the data measured by P-D technique has the resolution in frequency measurement,
$\delta f_{sig}$, determined by  $\delta f_{sig} = f_{sig}/(f_{samp}T_s)$, while that
using traditional FFT technique is $\delta f_{sig}= 1/T_s$, where $f_{samp}$ is the
sampling frequency ($\sim$ 1000\,Hz) and $T_s$ is the signal duration in time. For the
decay signal presented in Fig. \ref{fig1}, the fit to experimental data measured by the
P-D technique gives the resolution in frequency to be $\sim$ 30\,$\mu$Hz, while the
resolution of traditional FFT technique is $\sim$ 10\,mHz.
\begin{figure}[htb]
\begin{center}
\epsfig{figure=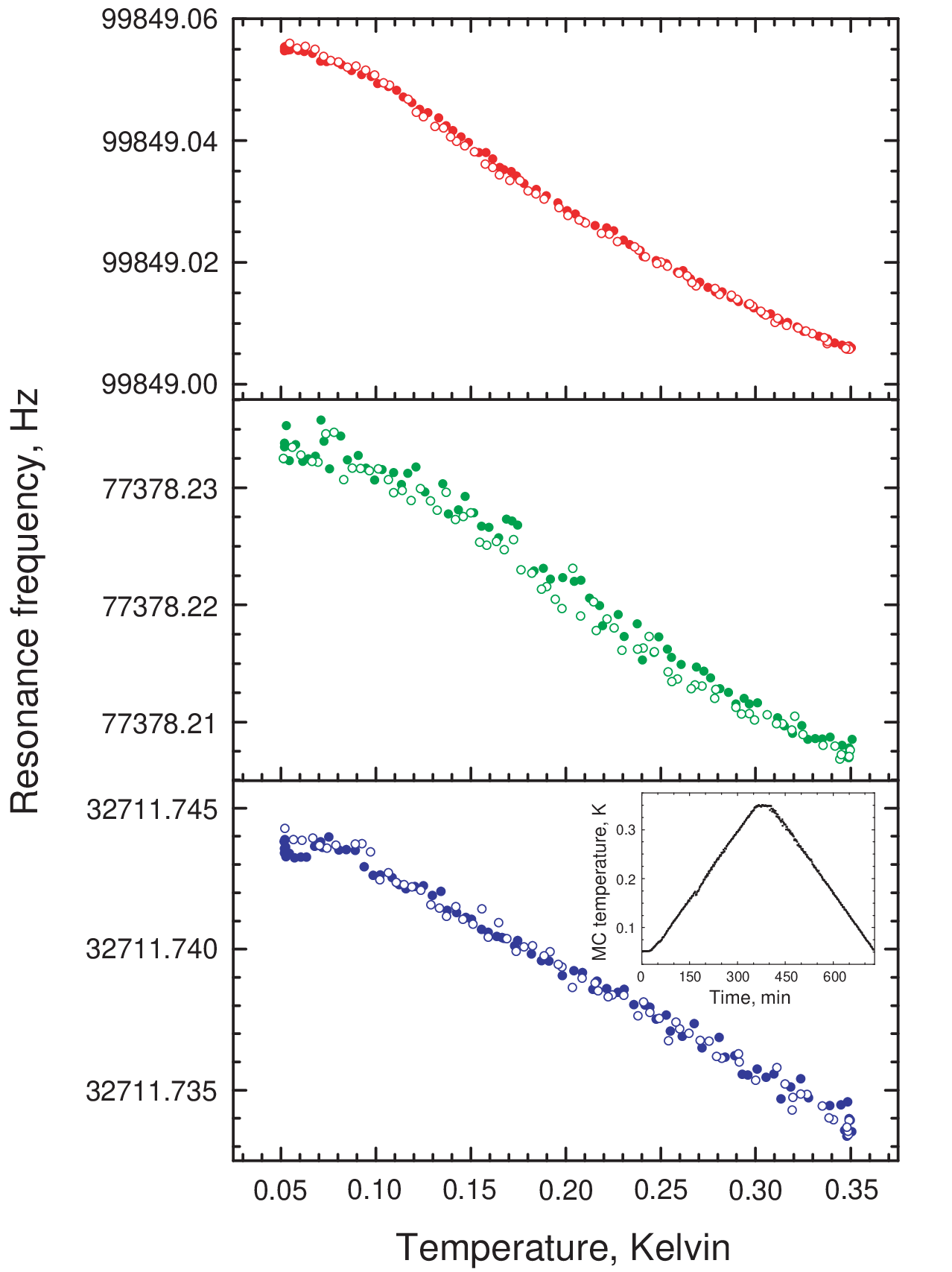,width=80mm}
\caption{(Colour online) Dependencies of the tuning forks'
resonance frequencies measured using pulsed-demodulation technique as a function of the mixing
chamber temperature during warming up (solid dots) and subsequent cooling down (empty dots) process.
Inset shows the time evolution of the mixing chamber temperature.} \label{fig2}
\end{center}
\end{figure}

Figure \ref{fig2} shows measured temperature dependencies of the tuning forks' resonance
frequencies during warming and cooling sweeps of the dilution refrigerator in temperature
range between 50\,mK and 350\,mK. Temperature of the mixing chamber was measured using
SQUID noise thermometer MFFT-1 provided by Magnicon and the temperature calibration was
cross-checked using home-made fixed-point device \cite{fpd}. The rate of the temperature
sweeps was 1\,mK per minute. We found that during warm up process the resonance
frequencies of the tuning forks are decreasing. At a constant temperature, the resonance
frequencies stay constant and on subsequent cooling process, the resonance frequencies
rise reproducibly again. It is worth to note that we were not able to measure the
temperature of the tuning forks themselves and presented measurements are related to
temperature of the mixing chamber. However, based on the reproducibility we assume that
the tuning forks were in thermal equilibrium with mixing chamber in temperature range
presented. Inset to Fig.~\ref{fig2} shows the time evolution of mixing chamber
temperature.  What could be a physical origin of the fine temperature dependence of
tuning forks' resonance frequencies? Below we present a simple phenomenological physical
model.

In contrast to a classical mechanical oscillator,  there are oscillating induced electric
dipoles $\mathbf{p}=Q \mathbf{d} = \alpha (T) \mathbf{E}_{loc}$ ($Q$ is the charge and
$\mathbf{d}$ is the distance between the charges that is proportional to deflection of
the ions from equilibrium position $\mathbf{x}$) in the tuning fork, and each of them
experiences a local electric field intensity $\mathbf{E}_{loc}$ produced by other
electric dipoles (and an initial voltage pulse). Once the electric dipole moment
$\mathbf{p}$ is formed its potential energy equals to  $- \mathbf{p}.\mathbf{E}_{loc}=
-\,p E_{loc}\cos (\theta)$. This energy has a tendency to orient the dipole $\mathbf{p}$
into direction of local field $\mathbf{E}_{loc}$, but on the other side, the thermal
energy $(k_B T)$ violates this dipole order. We presume that during "thermodynamic"
transition into high Q-value mode the potential energy of the van der Waals interaction
overwhelms the thermal energy and, similar to the dipole-dipole interaction in superfluid
$^3$He, this energy is always minimized i.e. it orients the dipole $\mathbf{p}$ into
direction of the electric field  $\mathbf{E}_{loc}$ and in such way provides the phase
coherence. The total potential energy of the dipole including the elastic energy
associated with lattice deformation
\begin{equation}
V_{pot} \approx \left( k + \frac{Q^2}{\alpha (T)}\right) x^2 \label{equ2}
\end{equation}
produces the restoring force $\mathbf{F}_{res} = - \nabla V_{pot}$, where $k$ is the
spring constant, $\alpha (T)$ is the temperature dependent polarizability and $x$ is the
deflection of the ion from equilibrium value.
 Polarizability of the quartz excited by the pulse can be expressed in form
\cite{blakemore}
\begin{equation}
\alpha (T) = \frac{3 \varepsilon_0}{N <\cos (\theta)>} = \frac{3 \varepsilon_0}{N
L(X)},\label{equ4}
\end{equation}
where $\varepsilon_0$ is the vacuum permittivity, $N$ is the density of dipoles and
$<\cos (\theta)>$ is the mean value of $\sum_i \cos(\theta)_i$, which is equal to the
Langevin function $<\cos (\theta)> = L(X) = \coth(X) - 1/X$ with $X = p\cdot E_{loc}/(k_B
T)$. Therefore, the frequency of the tuning fork oscillations $f_{tf}$ is determined by
two terms
\begin{equation}
f_{tf}(T)^2 = \frac{k}{4\pi^2\,m_{eff}} + \frac{Q^2 N L(X)}{12\pi^2\, \varepsilon_0 m_{eff}} = f_0^2 +
f_T^2(T).\label{equ5}
\end{equation}
While the first, constant term, in presented model, corresponds to the resonance
frequency determined by the elastic properties of the quartz, the second term reflects
the temperature contribution to the resonance frequency due to the rising stiffness of
the potential energy provided by the van der Waals interaction with decreasing
temperature.

\begin{figure}[htb]
\begin{center}
\epsfig{figure=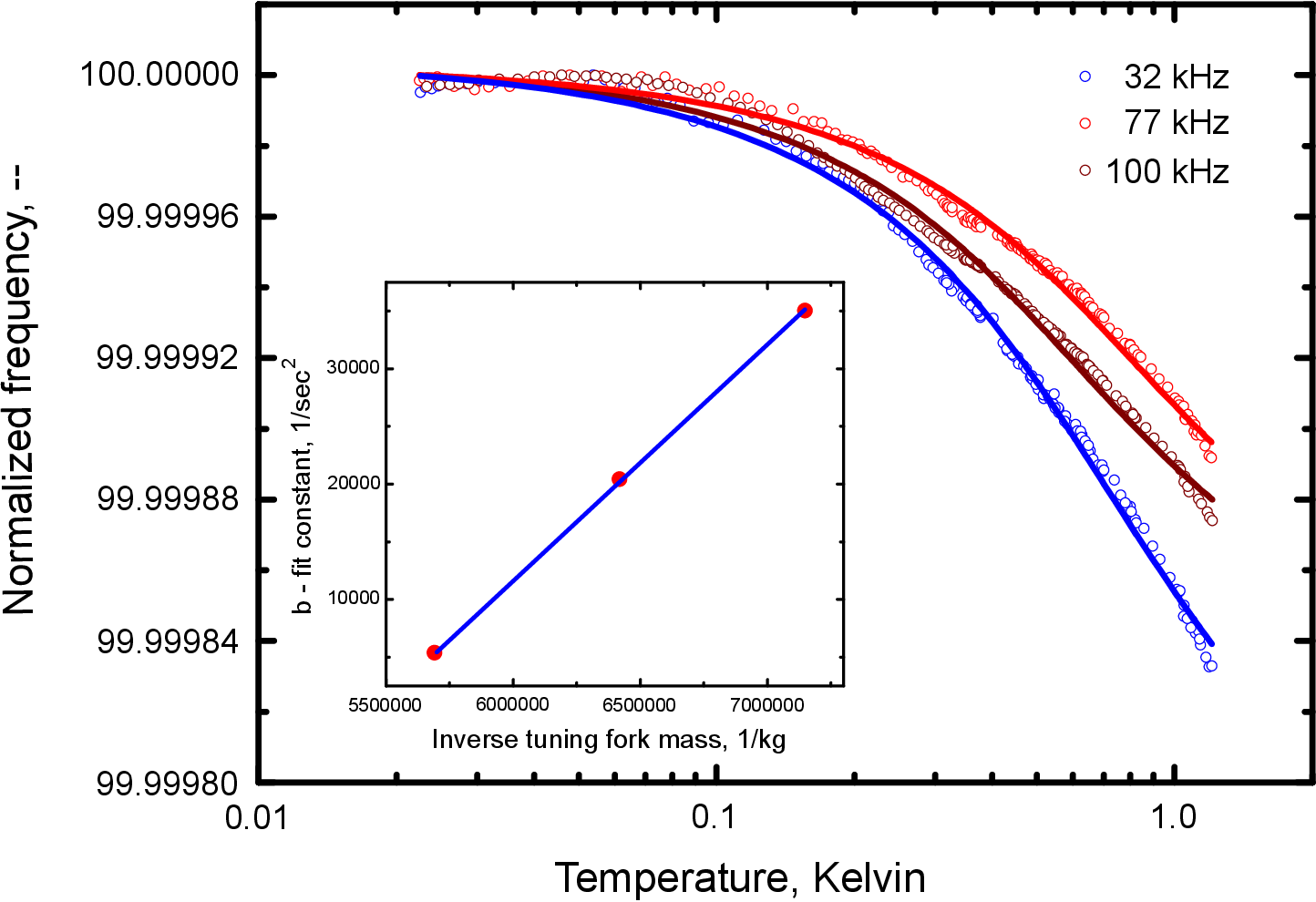,width=85mm} \caption{(Colour online) Temperature dependencies of the
normalized tuning forks' resonance frequencies. Solid lines  are the fits to experimental
data using equation $f_{tf} = \sqrt{a + b.L(T_c/T)}$. Inset shows the dependence of the $b\,$-constants
as a function of the inverse effective mass $(1/m_{eff})$. The line corresponds to
the linear fit to the data.} \label{fig3}
\end{center}
\end{figure}

Figure \ref{fig3} shows the temperature dependencies of the tuning forks' resonance
frequencies normalized to the maximal values, measured in temperature range from 25\,mK
to $\sim$ 1\,K. Solid lines represent the fits to experimental data using expression
$f_{tf} = \sqrt{a + b.L(T_c/T)}$ (see Eq. (\ref{equ5})), where $a = f_0^2$, $b$ is the
fitting constant characterizing the additional contribution to the stiffness due to van
der Waals interaction, $T_c = pE_{loc}/k_B$ is the critical temperature and $L(T_c/T)$ is
the above-mentioned Langevin function. Presented fits show a good qualitative agreement
with the experimental data. In fact, the values of critical temperatures $T_c$  for all
three forks are almost the same: they are in range between 1.1\,K and 1.5\,K, which
allows to estimate the energy of the van der Waals interaction to be of the order of
$\sim$ 2$\times$10$^{-23}$\,Joule ($\sim$ 0.12\,meV).

Inset to Fig. \ref{fig3} shows the dependence of the $b$-constants obtained from the fits
as a function of the effective mass $m_{eff}$. The $b$-constant is expressed as
\begin{equation}
b = \frac{Q^2 N}{12\pi^2\, \varepsilon_0 m_{eff}} = A \left(\frac{1}{m_{eff}}\right).
\end{equation}\label{equ6}
The values of the effective mass for individual tuning forks were calculated using
expression $m_{eff}=0.25 \emph{L}\emph{T} \emph{W} \rho_{SiO}$, where
$\emph{L},\emph{T},\emph{W}$ are the geometrical length, the thickness and the width of
the arms of individual forks, respectively, and $\rho_{SiO}$ is the mass density of the
$SiO_2$. Pre-factor $0.25$ counts for an effective displacement of the forks' arm (mass)
during oscillations. The calculated values of the  effective mass for individual forks
are as follow: 32\,kHz fork - $m_{eff}$= 1.7573E-7\,kg; 77\,kHz fork - $m_{eff}$=
1.5582E-7\,kg; 100\,kHz fork - $m_{eff}$= 1.3992E-7\,kg. A linear dependence of the
experimental values of  $b$-constants on $1/m_{eff}$ shown in inset of Fig. \ref{fig3}
supports presented phenomenological model.

\begin{figure}[htb]
\begin{center}
\epsfig{figure=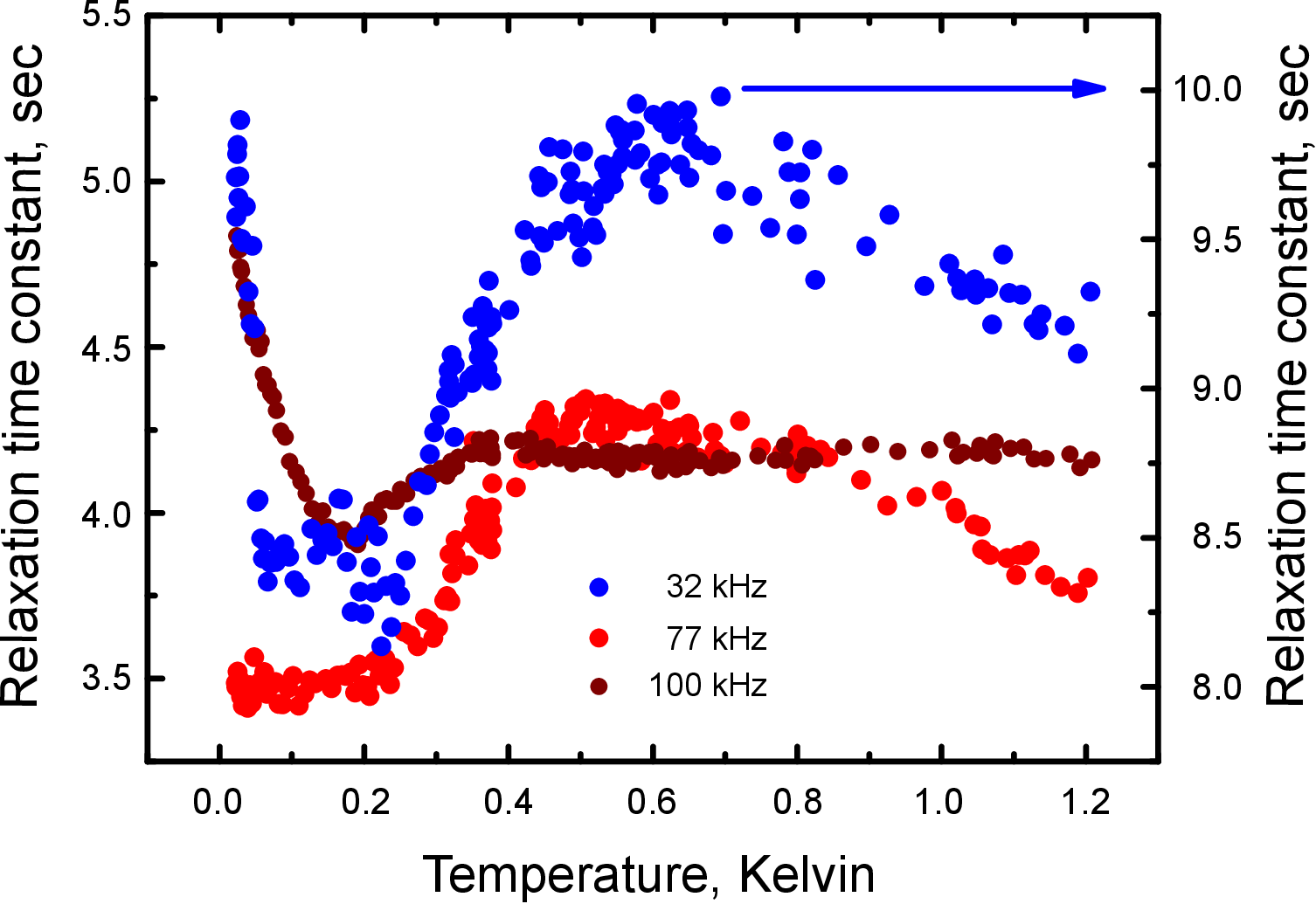,width=85mm} \caption{(Colour online) Temperature dependencies of the relaxation time constants $\tau$ determined from the
same free decay signals. We note that the relaxation time constant for 32\,kHz fork is related to right $y$-axis.} \label{fig4}
\end{center}
\end{figure}

Now, let us discuss the influence of intrinsic damping process on tuning forks' resonance
frequencies. As mentioned above, the intrinsic damping processes are reflected in the
relaxation time constant ($\gamma_D \sim 1/\tau (T)$). Figure \ref{fig4} shows measured
temperature dependencies of the relaxation time constants $\tau$ for individual tuning
forks. Dependencies presented in Fig. \ref{fig4} clearly demonstrate contradiction with
properties of the standard linear harmonic oscillators, where the resonance frequency
depends on damping ($\omega^2 = \omega^2_0 - \gamma_D^2/4$). Temperature dependencies of
the resonance frequencies for measured forks behave independently on damping process, and
this reveals  that dissipation mechanisms are acting rather near the quartz surface than
in its volume i.e. at the boundary, where the coherence is violated. Moreover,
temperature dependencies of relaxation time constants $\tau$ show ``Schottky-like
anomalies'' i.e. the temperature extrema, which one can attribute to a presence of
thermally activated dissipation mechanisms. As the surfaces of the individual quartz
tuning forks are  covered by the metal electrodes made of various elements (tin, silver,
etc.), Schottky barriers are formed on this interface due to difference in energy spectra
between metal electrodes and quartz \cite{tersoff}. These Schottky barriers can open
additional dissipation channels caused by thermally activated injection of the electric
charges from metal alloy to quartz. However, the physical nature of the dissipation
processes remain unclear up to now and an additional work needs to be done to elucidate
this problem.

\begin{figure}[htb]
\begin{center}
\epsfig{figure=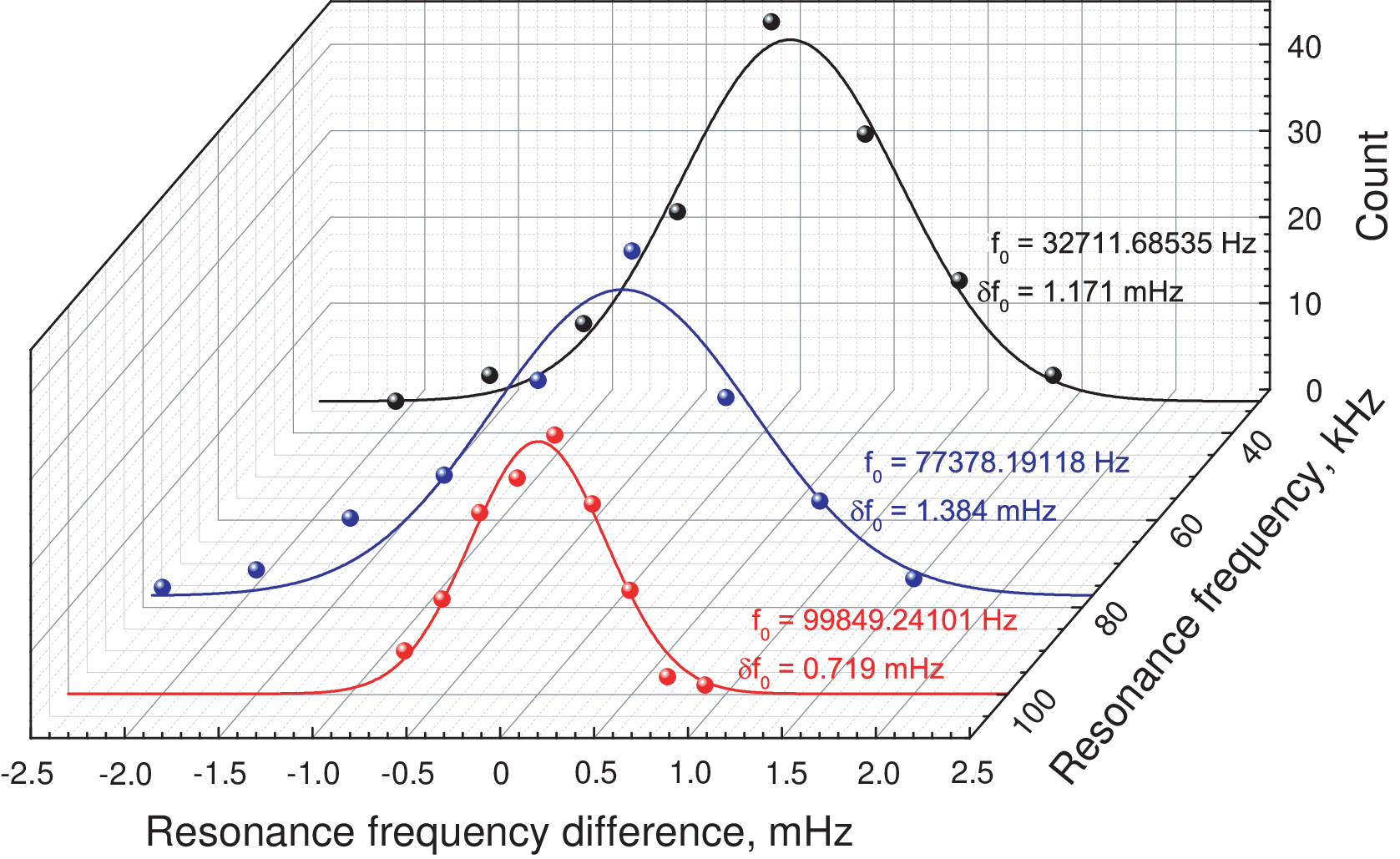,width=85mm} \caption{(Colour online) Distributions of the tuning forks'
resonance frequencies measured by the P-D technique at constant temperature of $\sim$ 8\,mK. Lines
show the Gaussian fits to experimental data.} \label{fig5}
\end{center}
\end{figure}

Observed temperature dependencies of the tuning forks' resonance frequencies and their
independence on the damping processes indicate that the system of oscillating dipoles in
quartz tuning forks preserves the phase rigidity i.e. the phase coherence due to van der
Waals interaction, thus revealing the characteristics of a coherent state of the
oscillating dipoles. In order to justify this assumption we performed several
measurements of the temperature stability of the resonance frequencies of individual
tuning forks at different but constant temperatures. Figure \ref{fig5} presents the
distributions of the tuning forks' resonance frequencies measured at base temperature of
the dilution refrigerator. In order to show all three dependencies, we plot the
difference from resonance frequency obtained from the Gaussian fits to measured data of
individual tuning forks. Presented dependencies demonstrate extremely well defined
resonance frequency for all tuning forks - for 32\,kHz fork: 32711.68535\,Hz $\pm$
32\,$\mu$Hz, for 77\,kHz fork: 77378.19118\,Hz $\pm$ 57\,$\mu$Hz and for 100\,kHz fork:
99849.24101\,Hz $\pm$ 19\,$\mu$Hz, with relatively narrow line-width $\delta f_0$ equal
to 1.171\,mHz for 32\,kHz fork, 1.384\,mHz for 77\,kHz fork and 0.719\,mHz for 100\,kHz
fork. Temperature of the mixing chamber was 8.2\,mK $\pm$ 80\,$\mu$K during measurements.
Taking a temperature derivative of the expression (\ref{equ5}), one can estimate a
frequency variation $\delta f_0$ with temperature changes $\delta T$. Calculated values
of the frequency variations $\delta f_0$ corresponding to the temperature changes $\delta
T \approx 80\, \mu$K for all forks are of the order of 10\,$\mu$Hz. Comparison of the
calculated values of $\delta f_0$ with measured data suggests that tuning forks
themselves are exposed to a heating effect or there are temperature gradients present
inside the forks leading to the frequency fluctuations, but sources of the heat or origin
of the temperature gradients remain unknown yet.

The measurements of the piezo-resonator's resonance frequencies at constant temperature,
i.e. the frequency stability were also analyzed in terms of the Allan deviation
\cite{allan,cleland}. The calculated Allan deviations of the resonance frequencies for
the individual tuning forks are: 1.0E-8 for 32\,kHz fork, 1.5E-8 for 77\,kHz fork, and
2.4E-9 for 100\,kHz fork. Considering that "the sample time" is of the order of several
tens of seconds and relatively low values of the resonance frequencies, the values of the
Allan deviations demonstrate very high frequency stability comparable with the lasers
stability.

In order to show the presence of spontaneous emergence of the phase coherence in the
system of oscillating dipoles in quartz tuning forks, we tried to excite the fork's
oscillations using incoherent white noise signal, similar to the experiment performed
with B-E condensate of magnons in superfluid $^3$He-B \cite{spinlaser}.

\begin{figure}[htb]
\begin{center}
\epsfig{figure=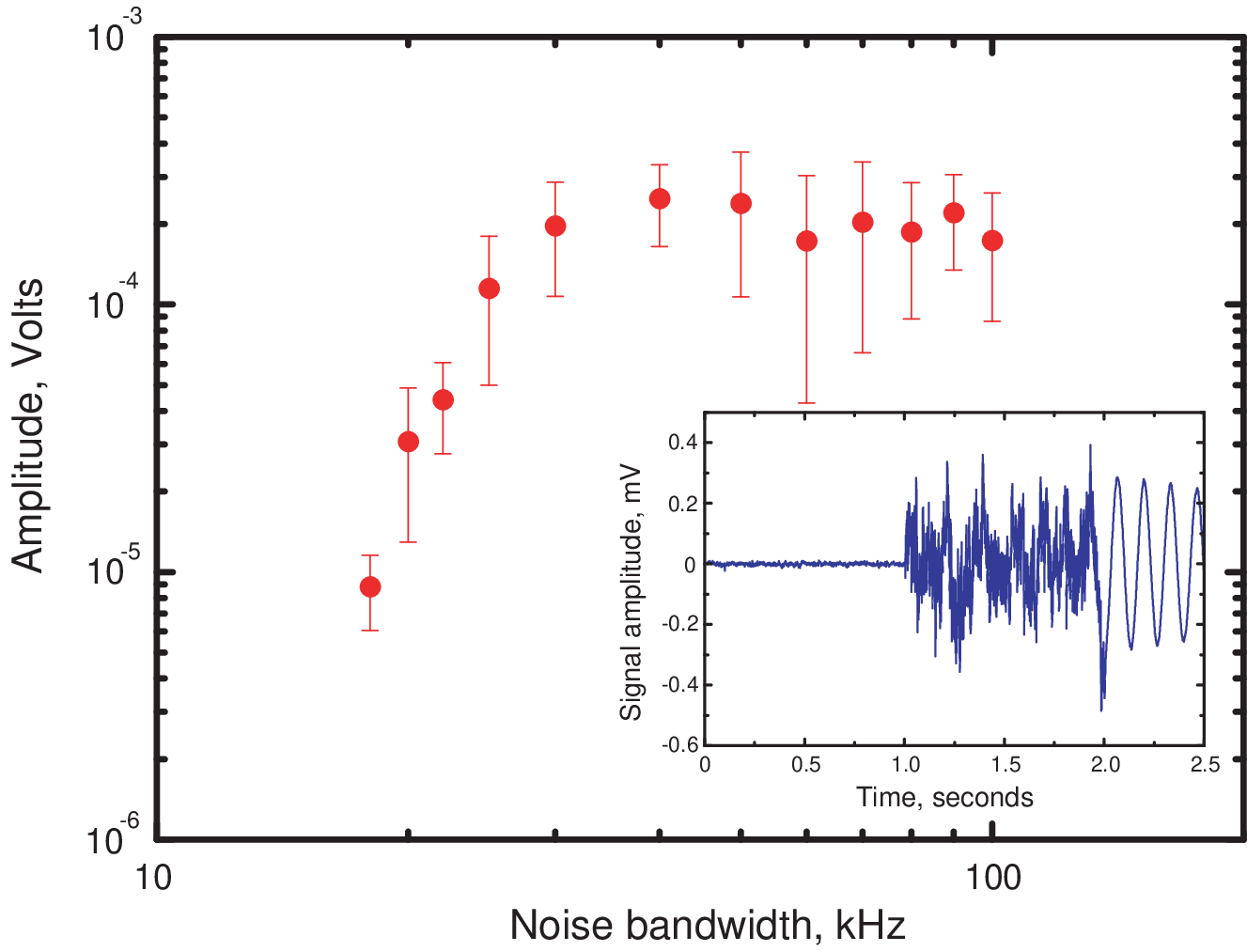,width=80mm} \caption{(Colour online) Dependence of the initial
signal amplitude as a function of frequency bandwidth of the noise signal applied to excite
oscillations of the tuning fork. Measurements were performed at temperature of 25\,mK. Inset shows
the birth  and formation of coherent signal from white noise excitation being applied for one
second after the idle state of the same duration.} \label{fig6}
\end{center}
\end{figure}

Noise measurements were performed by applying the white noise excitation with known
bandwidth for  1\,second period of time and amplitude of 5\,mV$_{rms}$ to the 32\,kHz
tuning fork. The inset to Fig. \ref{fig6} shows the dynamics of the birth and formation
of the coherent signal. Due to statistical nature of the noise signal, measurements of
the decay signals were repeated several times for given noise bandwidth. Figure
\ref{fig6} shows expected dependence of the signal initial amplitude as a function of
noise bandwidth: the tuning fork is excited by white noise, when the frequency bandwidth
of white noise includes the fork resonance frequency. It is worth to note that no decay
signals were detected, when the noise signal having the bandwidth less than 18\,kHz was
applied.

In conclusion -  we presented the experimental results on the spontaneous emergence of
the phase coherence in the system of oscillating electric dipoles in quartz
piezo-resonators caused by the van der Waals interaction. Spontaneous emergence of the
phase coherence in these systems is manifested via temperature-dependent, extremely
accurate tune-up of their resonance frequencies in 9th order with relative spectral
line-width $\delta f_0/f_0$ less than 3.10$^{-8}$ along with the very high frequency
stability characterized by the low values of the Allan deviations. These characteristics
open potential application of piezo-resonator as alternative time etalons. Moreover, we
showed that the application of an incoherent (noise) excitation signal leads to a
spontaneous formation of the phase coherent state and that dissipation processes do not
affect this phase coherent state (i.e. the resonance frequency of the system). All above
mentioned signatures are typical characteristics for a B-E condensate of excitations.
Whether this phenomenon is universal for the broad class of piezoelectric materials, and
what is the nature of the dissipation mechanisms, these are the questions that need to be
answered.

We wish to thank to Grigory Volovik for fruitful comments. This work was supported by
projects APVV-14-0605, VEGA 2/0157/14, EU project  ERDF-ITMS 26220120005 (Extrem-I),
European Microkelvin Platform (H2020 project 824109) and by the U.S. Steel Ko\v{s}ice.

\end{document}